\documentclass[conference]{IEEEtran}
\usepackage{amsmath}
\usepackage[pdftex]{graphicx}
\usepackage{amsmath,amssymb}
\usepackage{bm}
\usepackage{color}

\newtheorem{definition}{Definition}

\title{Gradient Flow Decoding for LDPC Codes}

\author{%
  \IEEEauthorblockN{
  		Tadashi Wadayama, Kensho Nakajima and Ayano Nakai-Kasai}
  \IEEEauthorblockA{\IEEEauthorrefmark{1}%
		Nagoya Institute of Technology,
		Gokiso, Nagoya, Aichi 466-8555, Japan,\\
 		wadayama@nitech.ac.jp}
}
\begin{document}
\maketitle

\begin{abstract}
The power consumption of the integrated circuit is becoming a significant burden, particularly for large-scale signal processing tasks requiring high throughput. The decoding process of LDPC codes is such a heavy signal processing task that demands power efficiency and higher decoding throughput. 
A promising approach to reducing both power and latency of a decoding process is to use an analog circuit instead of a digital circuit.
This paper investigates a continuous-time gradient flow-based approach for decoding LDPC codes, which employs a potential energy function similar to the objective function used in the gradient descent bit flipping (GDBF) algorithm. We experimentally demonstrate that the decoding performance of the gradient flow decoding 
is comparable to that of the multi-bit mode GDBF algorithm. 
Since an analog circuit of the gradient flow decoding requires only analog arithmetic operations and an integrator, 
future advancements in programmable analog integrated circuits may make practical implementation feasible.
\end{abstract}

\section{Introduction}

Recently, Moore’s law, which states that the number of transistors in an integrated circuit doubles every 24 months, is facing challenges due to the power consumption of the integrated circuit. The power consumption is becoming a significant burden to prevent handling large scale signal processing tasks requiring extremely high throughput such as multi-Gbits per second.

The decoding of low-density parity-check (LDPC) codes, first introduced by Gallager \cite{Gallager63}, is a heavy signal processing task requiring high throughput. In various fields, such as storage systems including solid-state drives and next-generation wireless systems beyond 5G or 6G, there is a demand for high-throughput  decoding algorithms. As a result, power efficiency and higher decoding throughput are becoming crucial topics for further research.

One promising approach to reducing power consumption is to reduce circuit size. Bit flipping (BF) decoding algorithms meet this demand, as they require fewer flip-flops in a circuit compared to conventional belief propagation (BP) decoding or min-sum decoding. Specifically, variations of the gradient descent bit flipping (GDBF) algorithm \cite{Wadayama10a} have been extensively studied for implementation.

An alternative approach to reducing both power and latency of a decoding process is to use an analog circuit instead of a digital circuit. 
An analog computer \cite{Ulmann20}\cite{Koppel2021} capable of simulating nonlinear ordinary differential equations (ODE)
is composed of analog adders, multipliers, integrators, and other nonlinear devices. 
In the field of optical computation, optical adders and multipliers have been realized and used for solving ODEs \cite{LU12}. A recent example is a neural network implemented in the optical domain \cite{Zhang2021}. Optical analog integrated circuits are currently under intensive study \cite{Capmany2020}, with reports showing that optical signal processing achieves both fast processing and extremely low power consumption. Although it is too early to evaluate their competitiveness against the digital circuits, we believe that there is plenty of room for further studies on the fully-analog implementation of decoding algorithms for LDPC codes. Furthermore, from a theoretical point of view, the development of continuous-time error-correcting dynamical systems appears to be an interesting topic to pursue \cite{Wadayama2022b}.

This paper investigates a {\em gradient flow}-based approach for decoding LDPC codes. The gradient flow dynamics is a continuous-time system that evolves the state to reduce a predefined potential energy. We will employ a potential energy function that is similar to the objective function used in the GDBF algorithm.

\section{Preliminaries}
\subsection{Notation}

The code length is denoted by $n$. 
A binary sparse matrix over $\mathbb{F}_2$,
$\bm H \in \mathbb{F}_2^{m \times n}$,
is a parity check matrix for an LDPC code. 
The binary linear code $\tilde C(\bm H)$ is defined by 
\begin{align}
\tilde C(\bm H) \equiv \{\bm{b}  \in \mathbb{F}_2^n \mid  \bm H \bm{b} = \bm{0} \}.		
\end{align}
A binary to bipolar transform $\beta: \mathbb{F}_2 \rightarrow \{1, -1\}$
defined as $\beta(0) \equiv 1$ and $\beta(1) \equiv -1$ transforms $\tilde C(\bm H)$
into the bipolar code defined by
\begin{align}
 C(\bm H) \equiv \{\beta(\bm{b}) \in \{1, -1\}^n \mid \bm{b}   \in \tilde C(\bm H) \}.	
\end{align}

The index sets $A(i)$ and $B(j)$ are defined as
\begin{align}
	A(i) &\equiv \{j \mid j \in [n], H_{i, j} = 1 \}, \quad i \in [m], \\
	B(j) &\equiv \{i \mid i \in [m], H_{i, j} = 1   \}, \quad j \in [n],
\end{align}
respectively, where $H_{i,j}$ denotes the $(i,j)$-element of $\bm H$.
The notation $[n]$ denotes the set of consecutive integers $\{1,2, \ldots, n\}$.
The multivariate Gaussian distribution with mean vector $\bm m$ 
and covariance $\bm \Sigma$ is denoted by ${\cal N}(\bm m, \bm \Sigma)$.

\subsection{Related works}

The focus of current research lies in optimization-based approaches. Several works have been developed in this category, with the celebrated work by Feldman \cite{Feldman03} on linear programming decoding providing a clear demonstration that the decoding problem can be viewed as an optimization problem.

Applications of interior point methods for solving a convex problem 
to decoding problems have been studied \cite{Vontobel08} \cite{Wadayama10b}.
A gradient descent formulation of a non-convex objective function 
leads to the GDBF algorithm \cite{Wadayama10a}.

Some variants of the GDBF algorithm, such as the noisy GDBF algorithm \cite{Sundararajan14}, have shown considerable improvement in decoding performance. Additionally, recent works have presented ADMM-based decoding algorithms for LDPC codes, 
including those by Zhang, Liu, and Wang \cite{Zhang13}\cite{Liu16}\cite{Liu16b}\cite{Wang20}, 
which have shown promising results. Although the advancement of these optimization-based decoding 
algorithms is significant, all of them are discrete-time algorithms. 

 Research on continuous-time dynamical systems for signal processing is a relatively new field and still in its early stages. However, it has shown promise in solving optimization problems and signal recovery tasks. For example, a MIMO MMSE detector based on continuous-time dynamics was proposed in \cite{Nakai2020}, and a sparse signal recovery dynamics was studied in \cite{wadayama2022a}. Further studies in this field, including the proposed work on LDPC decoding using gradient flow dynamics, can help to clarify the benefits and limitations of signal processing based on continuous dynamics.

\section{Potential energy function}

\subsection{Code potential energy function}

To define the gradient flow, we need an appropriate  potential energy function.
The {\em code potential energy function} for $C(\bm H)$ is a multivariate polynomial defined as
\begin{equation}
	h_{\alpha,\beta}(\bm{x}) \equiv \alpha \sum_{j = 1}^n (x_j^2 - 1)^2 
	+ \beta \sum_{i = 1}^m \left( \left(\prod_{j \in A(i)} x_j \right)  - 1 \right)^2, 
\end{equation}
where $\bm x = (x_1, \ldots, x_n)^T\in \mathbb{R}^n$.
The parameters $\alpha \in \mathbb{R}_+$ and $\beta \in \mathbb{R}_+$ control the 
strength of the constraints.
In the right-hand side of this equation, 
the first term represents the bipolar constraint for $\bm{x} \in \{+1, -1\}^n$,
and the second term corresponds to the parity constraint induced by $\bm H$, i.e.,
if $\bm x \in C(\bm H)$, we have 
$\left(\prod_{j \in A(i)} x_j \right) -1 = 0$
 for any $i \in [m]$. The code potential energy function
was first introduced in the work on proximal decoding algorithm \cite{Wadayama2021}.

Since the polynomial $h(\bm{x})$ has a sum-of-squares form, the polynomial can be 
regarded as a penalty function that yields a positive penalty value
for non-codeword vectors in $\mathbb{R}^n$.
The code potential energy $h_{\alpha,\beta}(\bm{x})$ is inspired by 
the non-convex parity constraint function used in the GDBF objective function \cite{Wadayama10a}.
The sum-of-squares form directly implies the most important property of $h_{\alpha,\beta}(\bm{x})$, i.e., 
the inequality 
$
		h_{\alpha,\beta}(\bm{x})  \ge 0
$
holds for any $\bm{x} \in \mathbb{R}^n$. The equality holds if and only if $\bm{x} \in C(\bm H)$.

\subsection{Gradient}
In the following discussion, we need the gradient of $h_{\alpha,\beta}(\bm{x})$.
The first-order derivative of $h_{\alpha,\beta}(\bm{x})$ with respect to $x_k (k \in [n])$ is given by
\begin{align} \nonumber
	\frac{\partial}{\partial x_k}h_{\alpha,\beta}(\bm{x}) 
	&= 4\alpha(x_k^2 - 1) x_k \\
	&+ 2\beta\sum_{i \in B(k) } \left( \left(\prod_{j \in A(i)} x_j \right)  - 1 \right) 
	\left(\prod_{j \in A(i) \backslash \{ k \} } x_j \right). 
\end{align}
The gradient $\nabla h_{\alpha,\beta}(\bm{x})$ is thus given by
\begin{equation}
	\nabla h_{\alpha,\beta}(\bm{x}) = \left(\frac{\partial}{\partial x_1}h_{\alpha,\beta}(\bm{x}), \ldots, \frac{\partial}{\partial x_n}h_{\alpha,\beta}(\bm{x})     \right)^T.
\end{equation}

The point $\bm{z} \in \mathbb{R}^n$ satisfying the equality 
$
 \nabla h_{\alpha,\beta}(\bm{z}) = \bm{0}
$
is a {\em stationary point} of $h_{\alpha,\beta}$.
For any codeword $\bm{x} \in C(\bm H)$, 
$x_k^2 = 1$ for any $k \in [n]$ and 
$
 \prod_{j \in A(i)} x_j = 1
$
holds for any $i \in [m]$. This means $\nabla h_{\alpha,\beta}(\bm x) = \bm{0}$,
i.e., a codeword vector is a stationary point of $h$.

\section{Gradient Flow Decoding}

\subsection{Gradient flow dynamics}

The gradient flow dynamics, also known as steepest descent dynamics, is a continuous-time dynamics defined by an ODE:
\begin{align}
	\frac{d\bm x(t)}{dt} = - \nabla F(\bm x(t)),
\end{align}
where $F:\mathbb{R}^n \rightarrow \mathbb{R}$ is a 
potential energy function. The system's state $\bm x:\mathbb{R} \rightarrow \mathbb{R}^n$
evolves to reduce the potential energy $F$ as the time variable $t$ increases.
This continuous dynamical system can be viewed as the continuous counterpart of the gradient descent method.
 If the potential function $F$ is strictly convex, 
the equilibrium point of the dynamics coincides with the minimum point of $F$.
Therefore, the gradient flow dynamics can be seen as a continuous-time 
minimization process of the potential energy.
If $F$ is a non-convex function, then
the gradient flow dynamics finds a stationary point of $F$ as $t \rightarrow \infty$.

\subsection{Gradient flow decoding}

We assume an AWGN channel in this paper.
A transmitter send a codeword $\bm s \in C(\bm H)$ and
a receiver obtain a received word 
\begin{align}
	\bm y = \bm s + \bm n,
\end{align}
where $\bm n \sim {\cal N}(\bm 0, \sigma^2\bm I)$.

Let $f$ be a potential energy function defined by 
\begin{align} \label{potential_energy}
	f(\bm x) \equiv \frac 1 2 \|\bm x - \bm y\|^2 + h_{\alpha,\beta}(\bm x).
\end{align}
The first term of the potential energy function can be regarded as
the negative log likelihood function for an AWGN channel.
The second term is a penalty term attracting $\bm x$ to a codeword of $C(\bm H)$.

The gradient flow decoding is simply a gradient flow dynamics based on 
the potential energy function $f$ that is given by
$
	{d\bm x(t)}/{dt} = - \nabla f(\bm x(t)).
$
This ODE can be rewritten as follows.
\begin{definition}[Gradient flow decoding]
The gradient flow (GF) decoding is defined by the ODE:
\begin{align}\label{GF_ODE}
	\frac{d\bm x}{dt} &= - (\bm x - \bm y + \nabla h_{\alpha,\beta}(\bm x)) \\
	\bm x(0) &= \bm x_0,
\end{align}
where $h_{\alpha,\beta}(\bm x)$ is the code potential energy 
and  $\bm x(0) = \bm x_0 \in \mathbb{R}^n$ is an initial point.
\end{definition}

In the above ODE, the notation $\bm{x}(t)$ is abbreviated to $\bm{x}$ for simplicity. If we have a suitably powerful analog circuit or analog computer, we can simulate the dynamics of this ODE. This simulation process is equivalent to a decoding process, where the equilibrium point represents the decoding result. A possible choice for the initial point is to set $\bm{x}_0 = \bm{0}$. Alternatively, for certain cases, a small-norm vector or a random vector could be used as the initial point.
\subsection{Euler method}
\label{Euler_sec}

To evaluate the decoding performance of the GF decoding (\ref{GF_ODE}), a numerical method is required to solve the ODE. The simplest numerical method 
for solving simultaneous nonlinear differential equations is the Euler method~\cite{Griffiths2010}. 
While the convergence order of the Euler method is lower than that of higher-order methods, 
it is straightforward to use and can provide sufficiently accurate solutions 
if the time interval is sufficiently fine discretized. 
Therefore, in the present study, we will use the Euler method to solve the ODE (\ref{GF_ODE}).

Suppose that we require to simulate the dynamics defined by the above ODE (\ref{GF_ODE})
in the time interval $0 \le t \le T$.
The interval is divided into $N$ bins where $N$ denotes the number of discretized intervals.
The discrete-time ticks
$
t_{k} = k \eta (k = 0, 1, \ldots, N)
$
defines the boundaries of the bins 
where the width of a bin is given by 
$
\eta \equiv T/N.
$
It should be noted that the choice of the bin width  $\eta$ 
is crucial in order to ensure the stability and the accuracy of the Euler method. 
By using the Euler method,
the ODE (\ref{GF_ODE}) can be approximated by 
\begin{align}
\bm x^{(k+1)} = \bm x^{(k)} - \eta \nabla f(\bm x^{(k)}) 	
\end{align}
for $k=0,1,2,\ldots, N-1$.
The initial condition of this recursion is $\bm x^{(0)} = \bm x_0$.

\subsection{Example of Decoding Process}

Here, we present a small example to illustrate a decoding process. 
Suppose that we have a repetition code of length 2, 
i.e., $C = {(+1, +1), (-1, -1) }$. 
The code potential energy function for $C$ is given by:
\begin{align}
	h_{1,1}(\bm{x}) = (x_1^2 - 1)^2 + (x_2^2 - 1)^2 + (x_1 x_2 - 1)^2,	
\end{align}
where the parameters $\alpha$ and $\beta$ are set to 1.
\begin{figure}[htbp]
\begin{center}
\includegraphics[scale=0.35]{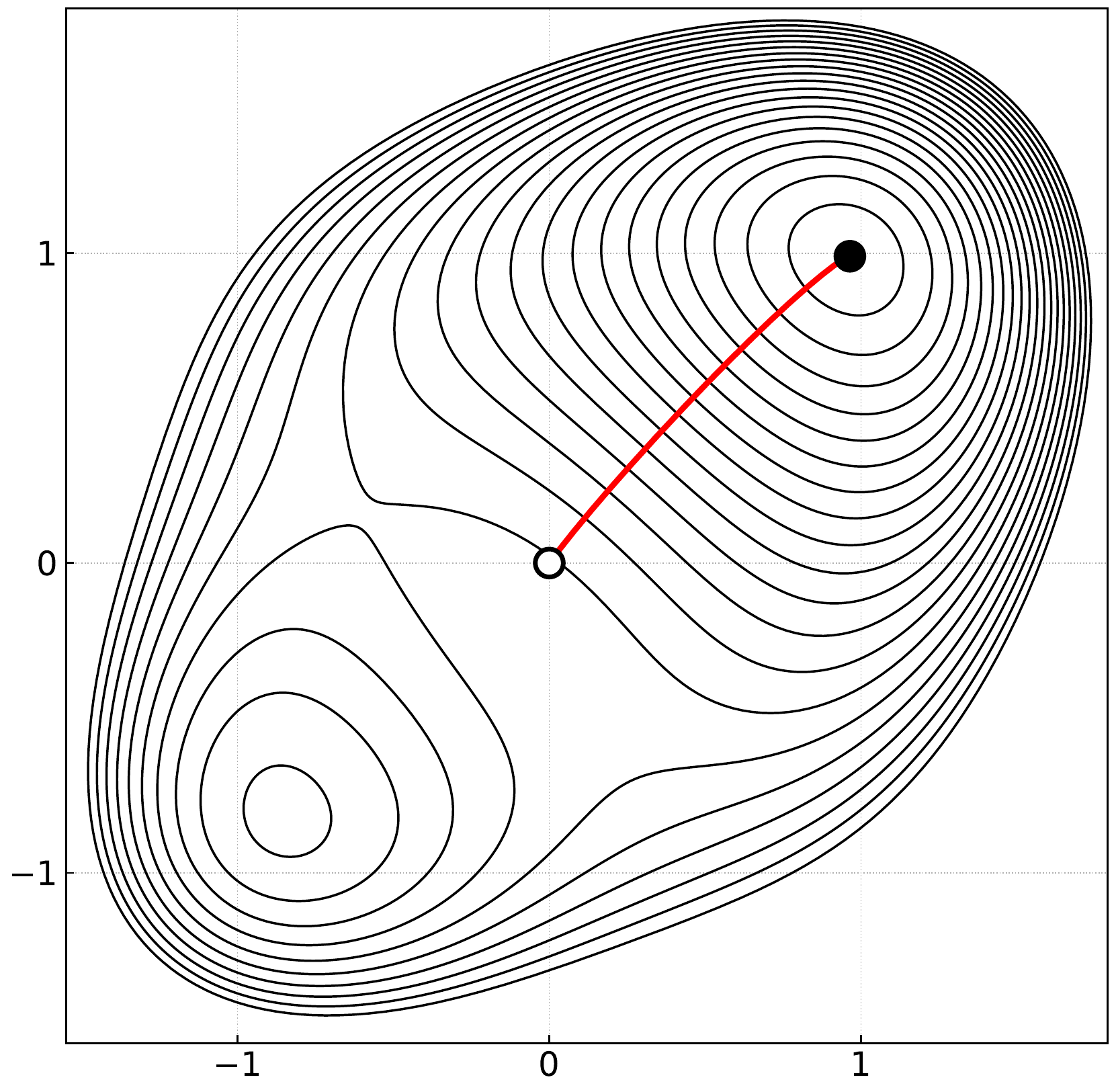}
\caption{Example of solution curve. The repetition code of length 2 is assumed.}
\label{fig:repetition}
\end{center}
\end{figure}

Assume that a transmitted word is $\bm x = (1,1)$ and 
that the corresponding received word is $\bm y = (0.6027, 0.8244)$.
In this case, the ODE (\ref{GF_ODE}) for the GF decoding becomes
\begin{align} \nonumber
&\left(
	\begin{matrix}
	\frac{dx_1}{dt} \\
	\frac{dx_2}{dt}
	\end{matrix}
\right)	 
= 
-\left(
	\begin{matrix}
	x_1 - 0.6027 + 4x_1(x_1^2 - 1) + 2 (x_1 x_2-1)x_2 \\
	x_2 - 0.8244 + 4x_2(x_2^2 - 1) + 2 (x_1 x_2-1)x_1
	\end{matrix}
\right).	
\end{align}

The initial condition for the ODE is set to $\bm x(0) = (0,0)$. Figure \ref{fig:repetition} shows the solution curve of the ODE, with the white small circle representing the initial point and the black small circle indicating the equilibrium point $(0.9642, 0.9901)$. The solution curve is obtained numerically using the Euler method. In Fig.~\ref{fig:repetition}, the contour curves of the potential energy function $f(\bm x)$ are also plotted. We can observe that the solution curve is orthogonal to the contour curves because the solution path follows the negative gradient vector field of the potential energy. This means that the curve is a steepest descent curve for $f$. By rounding the equilibrium point, we obtain $\hat{\bm x} = (1,1)$, which is the correct estimated word.

\section{Gradient Flow Dynamics}

This section focuses on observing the behavior of the gradient flow dynamics defined by (\ref{GF_ODE}), which provides insights into the decoding process of the GF decoding. The experiments in this section use a (3,6)-regular LDPC code with $n=204$ and $m=102$. The signal-to-noise ratio (SNR) is set to ${\sf SNR}=4$dB, and the standard deviation of the Gaussian noise $\sigma$ is given by 
\begin{align}
\sigma = \sqrt{(1/2)10^{-{\sf SNR}/10}/R},	
\end{align}
where $R$ is the design code rate. 
For example, $R$ equals $1/2$ for a (3,6)-regular LDPC code.
For the numerical results presented in this section, we used the Euler method with $N = 1000$ or $2000$. The parameters $\alpha$ and $\beta$ were both set to 1.

\subsection{Observation on potential energy}

If the state follows a gradient flow dynamics, the potential energy should decrease as the state evolves. To confirm the evolution of potential energy, we conducted an experiment as follows. For each trial, a random bipolar codeword $\bm s$ of a (3,6)-regular LDPC code was generated. The received word $\bm y$ was also randomly sampled  as $\bm y = \bm s + \bm n$, where $\bm n \sim {\cal N}(\bm 0, \sigma^2 \bm I)$. The received vector $\bm y$ was then provided to the gradient flow decoder. We conducted 50 trials in total.

\begin{figure}[htbp]
\begin{center}
\includegraphics[scale=0.35]{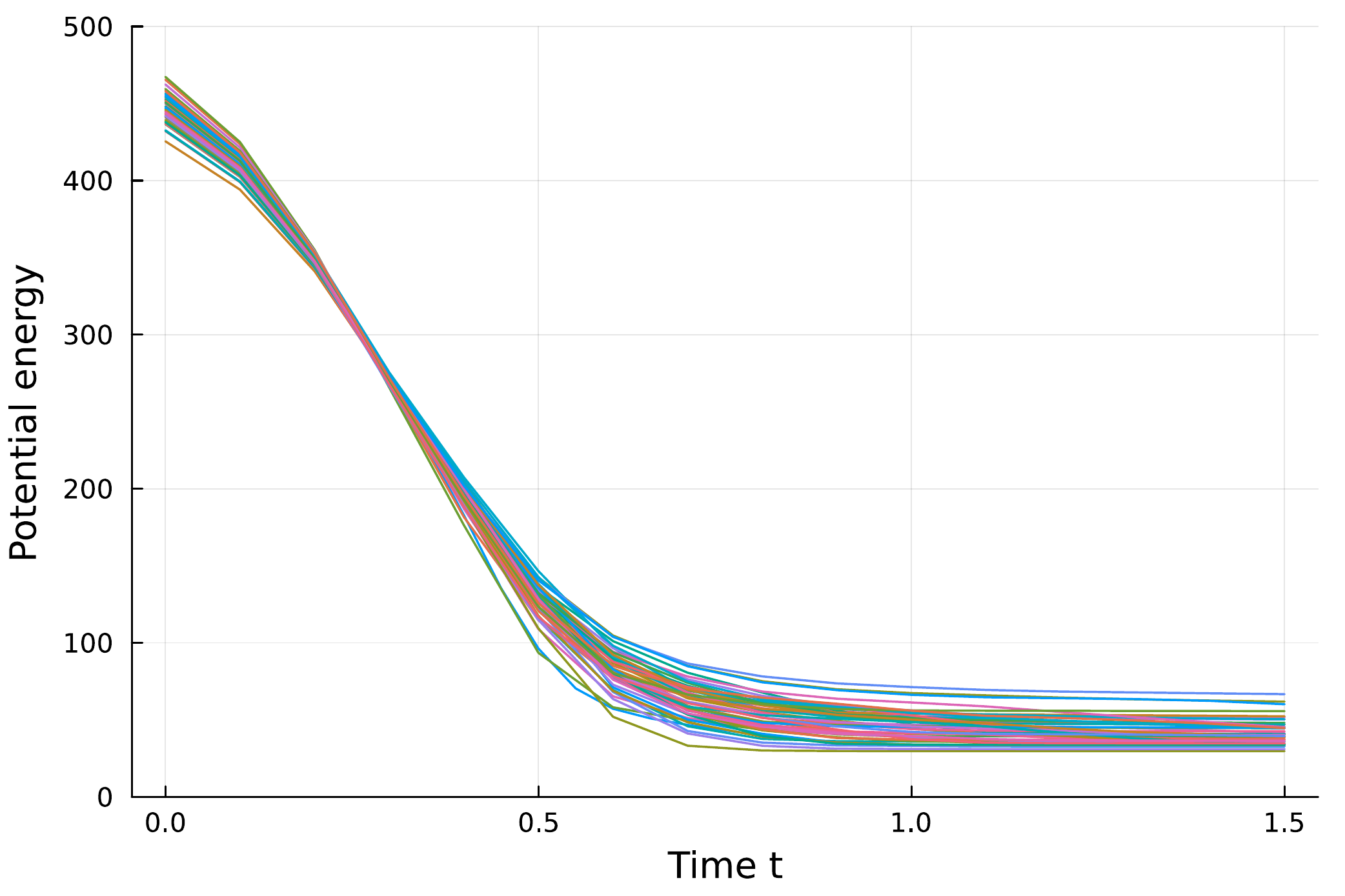}
\caption{Potential energy as a function of time $t$. (3,6)-regular LDPC codes of length 204 is used.}
\label{fig:potential_energy}
\end{center}
\end{figure}

Figure \ref{fig:potential_energy} plots the potential energy $f(\bm x(t))$ as a function of time $t$. As expected, we observed that the value of potential energy rapidly decreased. In the regime $t \geq 1$, the decrement of the potential energy was invisible, likely because the state vector was sufficiently close to the equilibrium point in such a regime.

\subsection{Decoding process}

\begin{figure}[htbp]
\begin{center}
\includegraphics[scale=0.35]{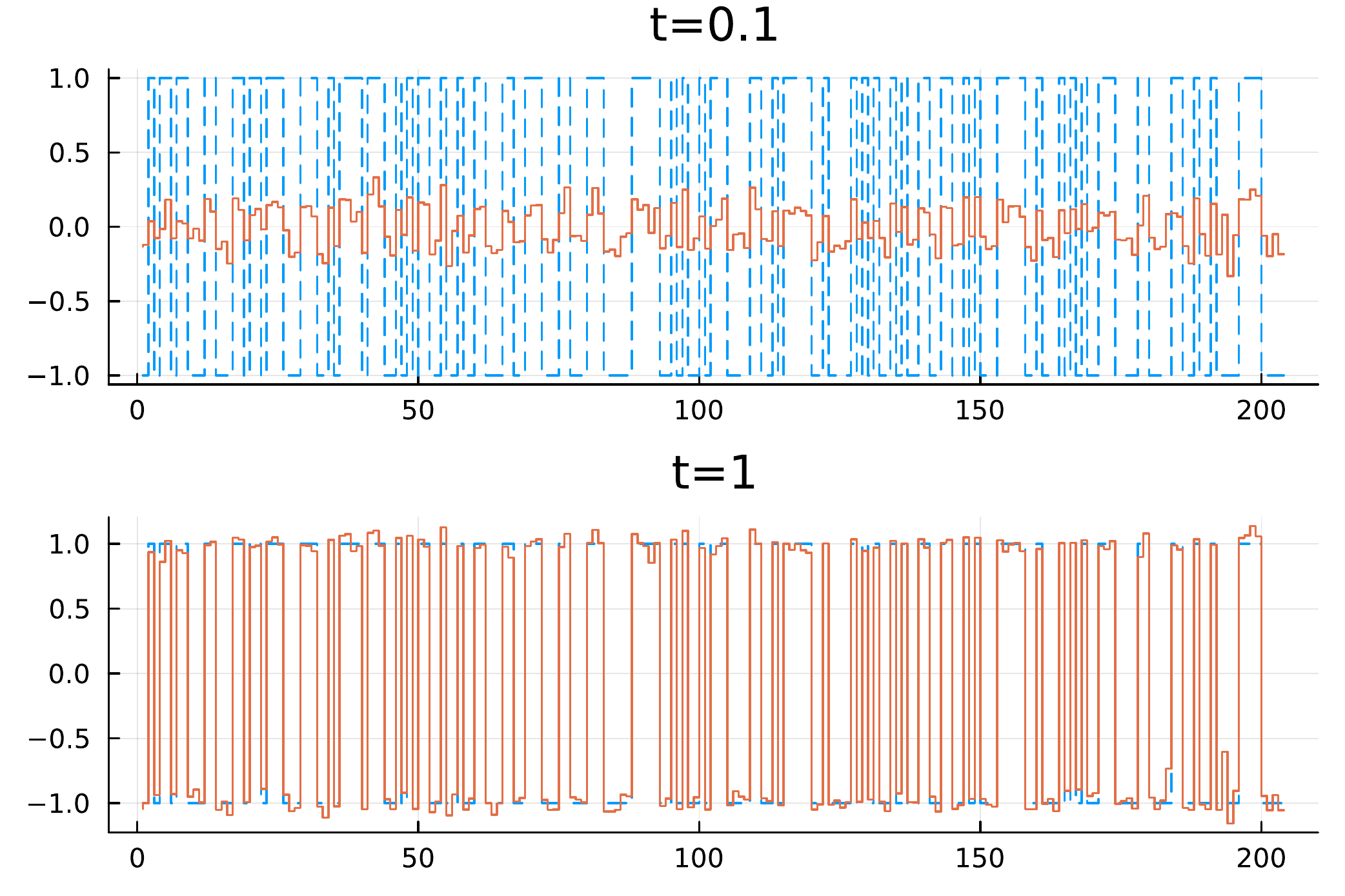}
\caption{Example of decoding process. The dashed line represents the original codeword. 
The solid line indicates the decoder output. ${\sf SNR} = 4$dB.}
\label{fig:dec_process}
\end{center}
\end{figure}

It may be informative to observe a decoding process of the GF decoding.
Figure \ref{fig:dec_process} presents a decoding process of the (3,6)-regular LDPC code.
The dashed line represents the transmitted word where the horizontal axis represents the 
index of a vector from $1$ to $n=204$. 
The solid line in Fig.~\ref{fig:dec_process} indicates the value of $x_i$ in $\bm x$.
The upper panel corresponds to time $t = 0.1$ and lower panel corresponds to 
time $t = 1$. At the beginning of the decoding process (upper panel), the magnitude 
of $x_i$ in the state vector $\bm x$ are fairly small. They are not sufficient to 
yield the final estimation. When $t = 1$ (lower panel), the values of $x_i$ are 
significantly close to the original vector. The state vector may be close to the equilibrium point.
This decoding result corresponds to successful decoding.

\subsection{Solution curves}

\begin{figure}[htbp]
\begin{center}
\includegraphics[scale=0.35]{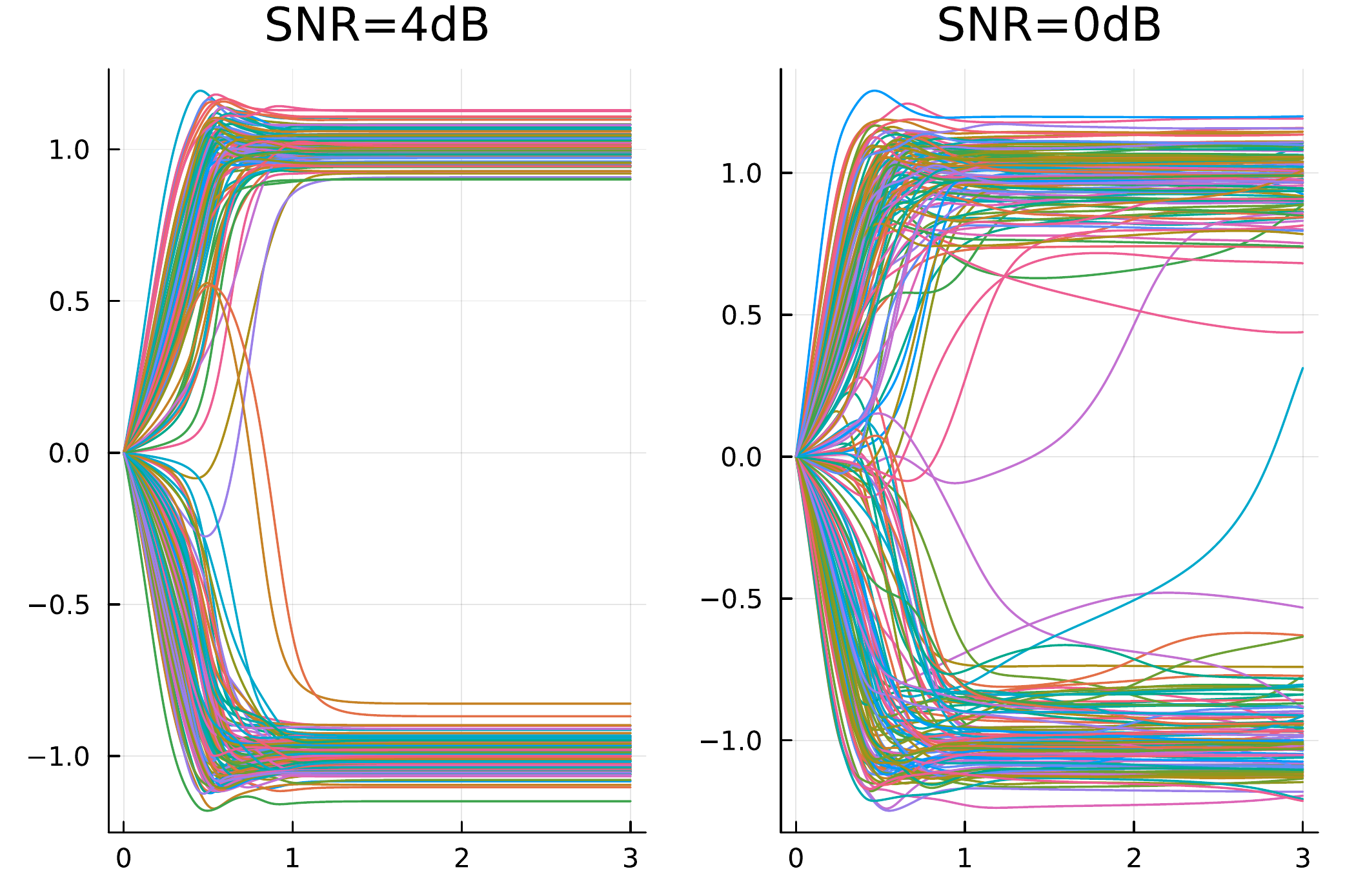}
\caption{Solution curves. (3,6)-regular LDPC codes of length 204.}
\label{fig:trajectory}
\end{center}
\end{figure}

Figure~\ref{fig:trajectory} displays the solution curves of each element in the state vector $\bm x$, where $x_i(t)$ (for $i \in [204]$) represents the solution of the ODE (\ref{GF_ODE}) plotted as a function of time $t$. Notably, all solution curves are smooth and continuous. The left panel illustrates the solution curves for ${\sf SNR} = 4$dB, where the equilibrium point is in proximity to a bipolar vector. Conversely, in the right panel (${ \sf SNR} = 0$dB), the solution curves are more widely dispersed.

\section{Decoder Architecture}

Our proposed scheme for decoding LDPC codes does not rely on any specific analog hardware technology for its implementation. We aim for a simple analog circuit to simulate the ODE (\ref{GF_ODE}) to facilitate easier implementation. 
In this section, we show a possible architecture of a GF decoder suitable for hardware implementation.
We assume that analog domain integrators exist and that basic real arithmetic operations, 
such as addition, subtraction, multiplication, and division, can be performed in the analog domain.

If $x_k \ne 0$ for any $k \in [n]$, then the partial derivative of $h_{\alpha,\beta}(\bm x)$
can be rewritten as
\begin{align} \nonumber
	\frac{\partial}{\partial x_k}h_{\alpha,\beta}(\bm{x}) 
	&= 4\alpha(x_k^2 - 1) x_k \\
	&+ \frac{2\beta}{x_k}\sum_{i \in B(k) } \left( \left(\prod_{j \in A(i)} x_j \right)  - 1 \right) 
	\left(\prod_{j \in A(i)} x_j \right).
\end{align}
Let $z_i (i \in [m])$ be intermediate variable defined by
$
	z_i \equiv \prod_{j \in A(i)} x_j.
$
We will introduce another intermediate variable $w_k (k \in [n])$ that is defined by 
\begin{align}
	w_k &\equiv 
	\sum_{i \in B(k) } \left(  z_i  - 1 \right) z_i 
	= \sum_{i \in B(k)} U(z_i),
\end{align}
where $U:\mathbb{R} \rightarrow \mathbb{R}$ is given by
\begin{align}
	U(z) \equiv (z-1)z.	
\end{align}
The function $V:\mathbb{R} \rightarrow \mathbb{R}$ defined by
\begin{align}
	V(w,x,y) \equiv -x + y -4\alpha (x^2-1)x - \frac{2\beta w}{x}	
\end{align}
is used to generate $\bm v = (v_1,v_2,\ldots, v_n)$ where 
\begin{align}
	v_k \equiv V(w_k, x_k, y_k).
\end{align}
By using the vector $\bm v$ obtained in the above way, we can rewrite
the ODE (\ref{GF_ODE}) by
$
	{d\bm x}/{dt} = \bm v.
$

\begin{figure}[htbp]
\begin{center}
\includegraphics[width=\columnwidth]{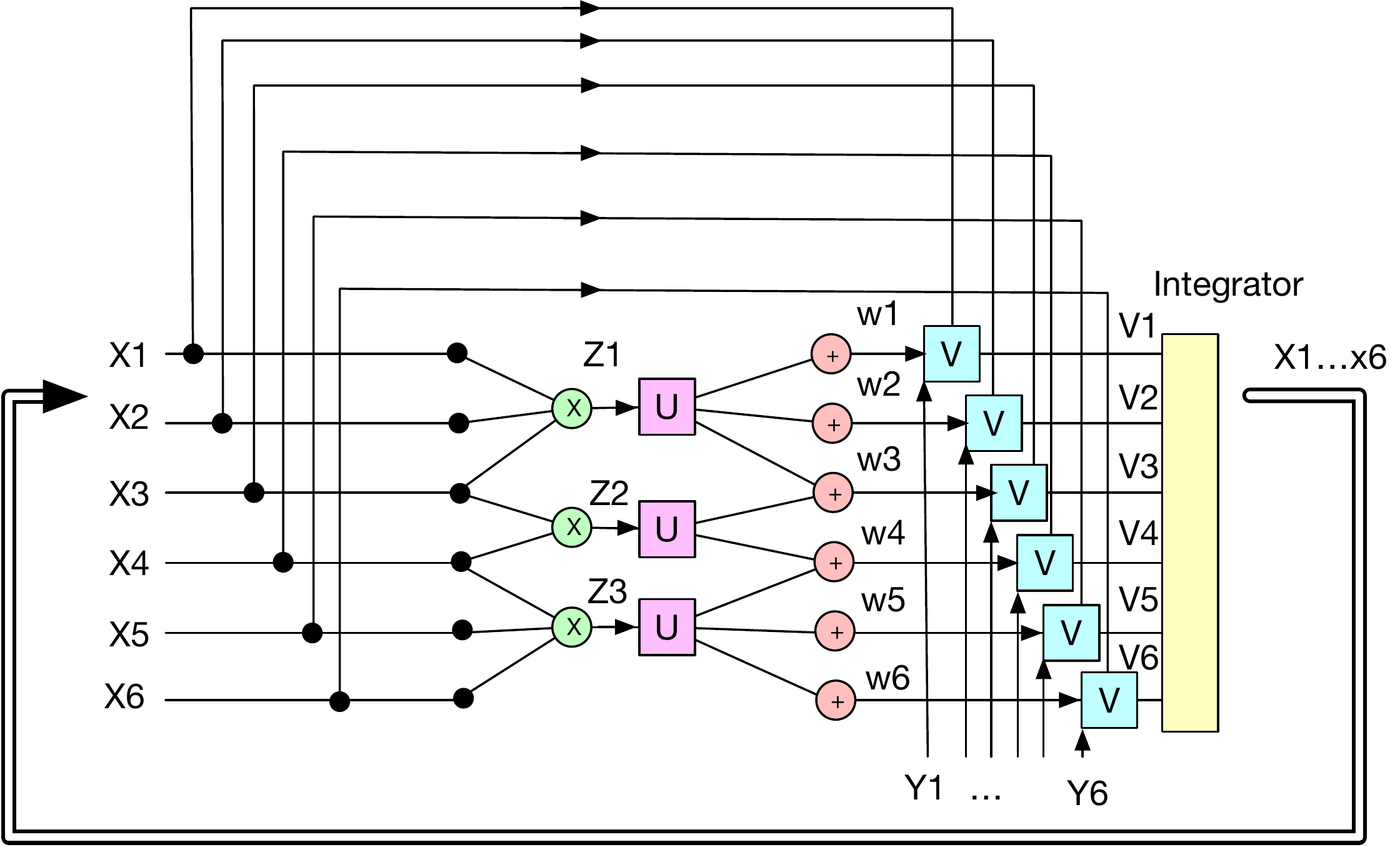}
\caption{Circuit diagram of gradient flow decoding.}
\label{fig:circuit}
\end{center}
\end{figure}

Figure \ref{fig:circuit} presents a configuration of a circuit having the gradient flow 
dynamics of (\ref{GF_ODE}). The parity check matrix 
\begin{align}
	\bm H 
	= 
	\left(
	\begin{matrix}
	1 & 1 & 1 & 0 & 0 & 0 \\
	0 & 0 & 1 & 1 & 0 & 0 \\
	0 & 0 & 0 & 1 & 1 & 1		
	\end{matrix}
	\right)
\end{align}
is assumed in this example.
The vector $\bm x = (x_1,x_2,\ldots, x_n)^T$ is entered to the 
circuit from the left. According to the Tanner graph of the parity check matrix $\bm H$, 
the intermediate variables $\bm z$ and $\bm w$ are evaluated.
Applying the function $V$ to $\bm w$, we finally obtain 
$
\bm v = - (\bm x - \bm y + \nabla h_{\alpha,\beta}(\bm x)). 	
$
By integrating the vector $\bm v$ with respect to time $t$, we have $\bm x$ and the vector 
is fed back to the input as in Fig.~\ref{fig:circuit}.
An analog domain integrator can converts $\dot{\bm x}$ to $\bm x$.

There are several remarks on the the analog circuit depicted in Fig.~\ref{fig:circuit}. Notably, the circuit lacks a component for initializing the value of $\bm x$ at the beginning. Given that the function $V(w,x,y)$ involves division by $x$, an initial point of $\bm 0$ is not feasible. An alternative solution is to use an initial vector of $\bm x_0 = \delta \bm y$ where $\delta$ is a sufficiently small real number.

\section{BER Performance}

We evaluated the bit error rate (BER) of GF decoding through computer simulations on LDPC codes with design rate 1/2, the $(3,6)$-regular LDPC codes (96.33.964, 204.33.484, PEGReg252x504, PEGReg504x1008) \cite{MacKay}. We used BP decoding as the baseline and set the maximum iteration of BP to 100. The parameter setting 
of GF decoding is summarized in Table \ref{parameters}.

\begin{figure}[htbp]
\begin{center}
\includegraphics[width=\columnwidth]{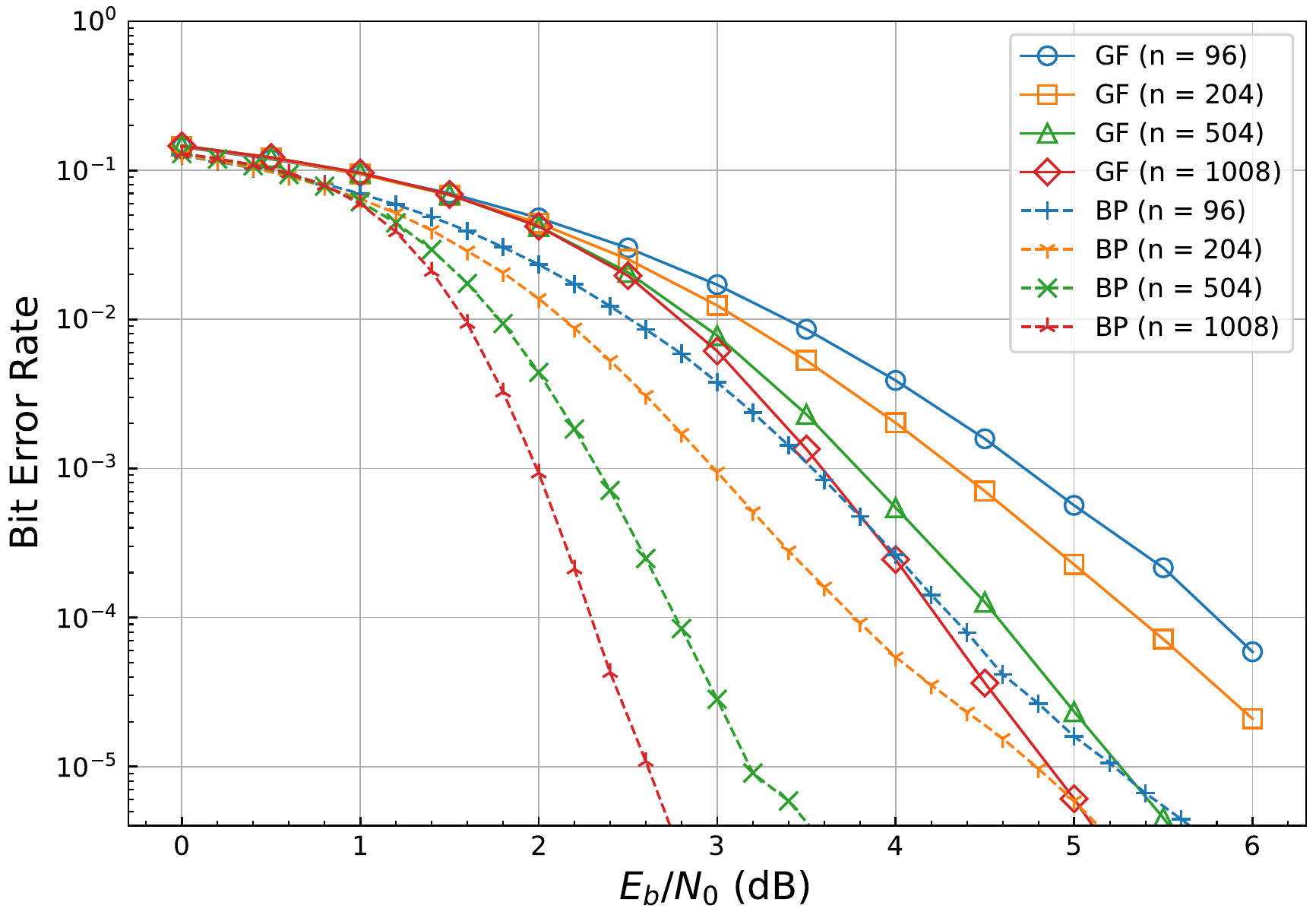}
\caption{Bit error rate of the GF decoding.}
\label{fig:BER}
\end{center}
\end{figure}

Figure \ref{fig:BER} displays the BER performance of the proposed GF decoding. Compared to BP, the GF decoding has a BER performance that is approximately 2dB worse. Notably, for PEGReg504x1008, the GF decoding performance is almost on par with the multi-mode GDBF algorithm using 100 iterations (\cite{Wadayama10a}, Fig.~3). Overall, GF decoding's BER performance is comparable to that of bit flip-type decoding algorithms.

\begin{table}
\caption{Parameter setting of GF decoding.}
\label{parameters}
\centering{
\begin{tabular}{ll}
	\hline
	\hline
Potential energy parameters &	$\alpha=1, \beta=2$ \\
Parameters of Euler method   & $T = 10, N = 1000$ \\
Sampling time & $t = 10$ \\
Encoding & Uniformly random codeword \\
\hline
\end{tabular}}
\end{table}

\section{Concluding summary}

We presented continuous-time dynamical systems for decoding LDPC codes. Specifically, the GF decoding is based on the gradient flow dynamics of the potential energy function (\ref{potential_energy}). 
We demonstrated that the decoding performance of the GF decoding is comparable to that of the multi-bit mode GDBF algorithm~\cite{Wadayama10a}. 
Advancements in programmable analog integrated circuits may make practical implementation feasible in the near future.

\section*{Acknowledgment}
This work was supported by JSPS KAKENHI Grant Number JP22H00514.

\bibliographystyle{IEEEtran}
\bibliography{gfd}

\end{document}